\documentclass[conference]{IEEEtran}
\IEEEoverridecommandlockouts
\usepackage{amsmath,amssymb,amsfonts}
\usepackage{amsthm}
\usepackage{textcomp}
\usepackage{xcolor}
\usepackage{hyperref}  % hyperlinks
\usepackage{url}  % simple URL typesetting
\usepackage{booktabs}  % professional-quality tables
\usepackage{nicefrac}  % compact symbols for 1/2, etc.
\usepackage{microtype} % microtypography
\usepackage{lipsum}
\usepackage{algorithm, algorithmic}
\usepackage{mathtools,nccmath}
\usepackage{array}
\usepackage{color,xcolor}
\usepackage{fancyhdr}
\usepackage{graphicx}
\usepackage{epstopdf}

\def\BibTeX{{\rm B\kern-.05em{\sc i\kern-.025em b}\kern-.08em
T\kern-.1667em\lower.7ex\hbox{E}\kern-.125emX}}
\begin{document}

\title{Distributed Clustering in the Anonymized Space with Local Differential Privacy}

\author{\IEEEauthorblockN{Lin Sun}
\IEEEauthorblockA{\textit{School of Software} \\
\textit{Tsinghua University}\\
Beijing, China \\
sunl16@mails.tsinghua.edu.cn}
\and
\IEEEauthorblockN{Jun Zhao}
\IEEEauthorblockA{\textit{School of Computer Science and Engineering} \\
\textit{Nanyang Technological University}\\
Singapore\\
junzhao@ntu.edu.sg}
\and
\IEEEauthorblockN{Xiaojun Ye}
\IEEEauthorblockA{\textit{School of Software} \\
\textit{Tsinghua University}\\
Beijing, China \\
yexj@mail.tsinghua.edu.cn}
}

\maketitle
\thispagestyle{plain} \pagestyle{plain}

\begin{abstract}
    Clustering and analyzing on collected data can improve user experiences and quality of services in big data, IoT applications. However, directly releasing original data brings potential privacy concerns, which raises challenges and opportunities for privacy-preserving clustering. In this paper, we study the problem of non-interactive clustering in distributed setting under the framework of local differential privacy. We first extend the Bit Vector, a novel anonymization mechanism to be functionality-capable and privacy-preserving. Based on the modified encoding mechanism, we propose kCluster algorithm that can be used for clustering in the anonymized space. We show the modified encoding mechanism can be easily implemented in existing clustering algorithms that only rely on distance information, such as DBSCAN. Theoretical analysis and experimental results validate the effectiveness of the proposed schemes.
\end{abstract}


\newtheorem{theorem}{Theorem}
\newtheorem{corollary}{Corollary}[theorem]
\newtheorem{lemma}[theorem]{Lemma}
\newtheorem{definition}{Definition}

\section{Introduction}

    Clustering is one of the most frequently-used methods in data-driven applications such as data mining, machine learning, computer vision, pattern recognition, recommender system and so on~\cite{zhang2014robust,mcsherry2009differentially}. Datasets could be divided into several classes by different clustering algorithms such as DBSCAN and k-means. Instances in the same classes have potential similarities, which are useful in data analysis. Taking user type analytics in electronic commerce applications as an example, clustering analysis on existing users with their characteristics and behaviors can be used in classifying new users to provide better shopping services.

    With a variety of data being collected and analyzed, the underlying privacy leakage of clustering must be stressed. However, most current clustering approaches are not privacy-preserving when designed. As show in \cite{wang2017cluster,hua2015differentially}, knowing the cluster results, one's precise position might be revealed in trajectory clustering with k-means clustering algorithm. Under such circumstances, how to control individual's privacy loss has became a substantial problem in big data analysis.

    Without loss of generality, the private data clustering can be classified into two approaches, the interactive and the non-interactive approaches. The interactive modes often follow these rules: a query function with its sensitivity is analyzed first, then noises are added to answers to these queries. In non-interactive settings, a synopsis of the input dataset are generalized and released for data analytics. To the best of our knowledge, most work focus on clustering in the interactive mode. \cite{su2016differentially} studies the trade-off between interactive vs. non-interactive approaches and proposes a clustering approach that combines both interactive and non-interactive. As far as we know, few work has been done for non-interactive clustering with local differential privacy.

    In the background of big data environment, data required in professional fields are usually dispersed in various data sources. Meanwhile, data are usually sparse, noisy and incomplete when collected~\cite{vatsalan2017privacy}. The performance of clustering on sparse and incomplete individual dataset will dramatically decline in those situations. Gathering distributed data can improve clustering performance, for example, by crowdsourcing~\cite{mazumdar2016clustering}. With IoT and cloud platform, data can be easily aggregated. However, collecting and analyzing data that supports complicated analyzing functions are hard to deploy. There are two main challenges:

    \textbf{A) How to control the underlying privacy leakage by data releasing.} Privacy is an important issue in data sharing, especially when those data are personal-related. Differential privacy is a strong privacy standard for privacy protection, and has significance in data releasing. For example, \cite{mohammed2014secure} uses taxonomy tree to publish data for vertically partitioned data among two parties and shows that the integrated data preserves utilities in classification tasks. However, applying differential privacy with exponential mechanism in choosing attribute does not anonymize source data, thus, data stays generalized which potentially leads to privacy leakage. To overcome this, we apply anonymization mechanism and then use local differential privacy. 

    \textbf{B) How to cluster on the anonymized dataset in distributed setting.} Currently, many methods have been proposed for privacy preserving clustering in the interactive mode~\cite{dai2016privacy}. These methods fail when it comes to the non interactive application areas. From the other perspective, many anonymization algorithms have been proposed. Currently, these mechanisms only allow statistical analysis such as mean and frequency estimation, thus can not be used for complicated analyzing. Retrieving necessary information in the anonymous space is needed.

\begin{figure}[t]
    \center
    \includegraphics[width=0.85\linewidth]{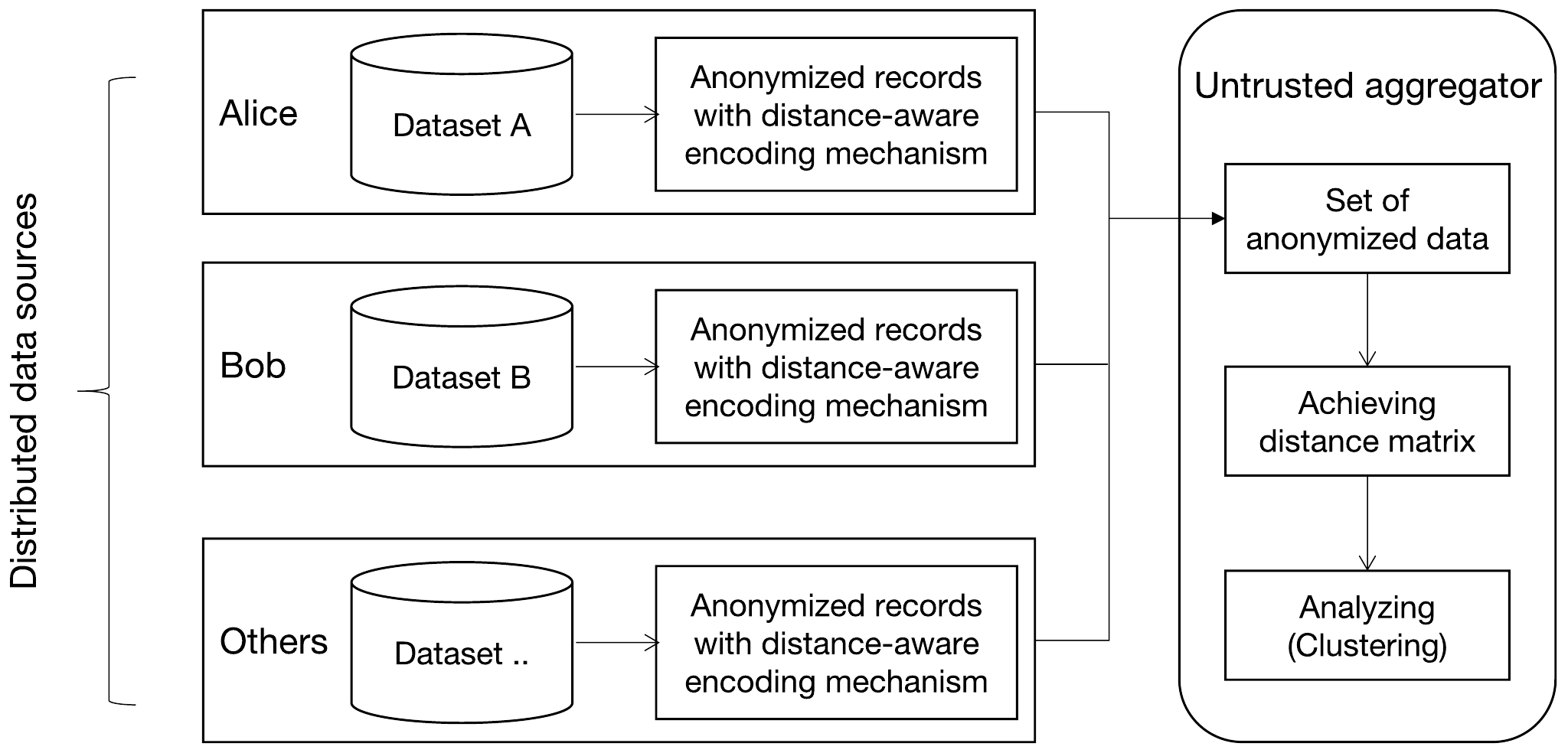}
    \caption{Framework of privacy-preserving clustering in distributed environments.}
    \label{fig:the framework of ppc}
    \end{figure}

    With these two considerations, a privacy-preserving encoding mechanism that supports clustering in the anonymous space is in demand. In this paper, we first use an encoding mechanism that embeds source data into anonymous Hamming space to eliminate semantic information. Then we add noise to provide indistinguishabilities based on local differential privacy. The anonymized data can then be released and consolidated. What's more, we show that the released dataset from multiple sources can be consolidated for clustering with the distance information retrieved from anonymous space. In general, the collecting and analyzing process is shown in Figure~\ref{fig:the framework of ppc}. There are three main steps:

    % 大数据下的隐私问题，引入差分隐私
    In the age of big data, personal-related data from user's side is routinely collected an which stands out the differential privacy~\cite{dwork2006calibrating, dwork2014algorithmic}.

    Firstly, all data custodians need to agree on the configurations parameters. In encoding, each data custodian embeds their data into anonymized space locally with compromised parameters. Then data are centralized to a data aggregator. At last, the anonymized data is analyzed, which contains distance matrix achieving and clustering. The encoding step should be privacy-preserving. Thus the privacy leakage is controlled during the whole process. For clustering utilities, distance information should be preserved in the anonymized space.

    We use the Bit Vector (BV) as basic encoding mechanism because of its distance aware property. However the localization of BV limits the clustering utilities and privacy-guarantee level. We solve this problem with a modification of BV mechanism. Then a clustering algorithm with only distance information is proposed. The contributions can be summarized as follows: 

    \begin{itemize}[leftmargin=*]
        \item We enroll the capabilities of Bit Vector mechanism by discovering distance consistence property in the anonymized space to make it suitable for whole range distance estimation.
        \item We expand the BV mechanism to be $(\epsilon, \delta)$- locally differentially private, which provides strict privacy guarantees for data sharing and analyzing.
        \item We show that the refined mechanism can be used for both horizontally and vertically partitioned data. Typically, for vertically partitioned data, we design the decomposition method that has lower estimation error.
        \item We show that the refined encoding mechanism can be used for clustering and can be easily used with existing methods.
    \end{itemize}

    The rest of this paper is organized as follows. In Section~\ref{section: related work}, related work and some preliminaries are presented. Then in Section~\ref{section: Towards functionalities: Bit Vector Mechanism}, we propose a distance consistence algorithm for whole range distance estimation and extend the BV mechanism to be $(\epsilon, \delta)$-differentially private. Based on the differentially private encoding mechanism, clustering algorithm on anonymous data are delineated in Section~\ref{section:clustering_with_anonymous_data}. We analysis experimental results in Section~\ref{section: experiments}. At last, in Section~\ref{section:conclusion}, we conclude this paper and discuss future work. Besides, the privacy analysis and limitations is presented in the Appendix.

\section{Related Work}
\label{section: related work}
    In this section, we briefly introduce the distance-aware encoding mechanism and the notion of differential privacy.

\subsection{Distance-aware Encoding Mechanism}
    The distance-aware encoding schemes try to embed source data into another space that preserves initial distance. For example, the bloom filters, together with N-grams are frequently used for string encoding as a solution in record linkage. The Euclidean distance is also used in a variety of application areas. Recently, some encoding mechanisms have been proposed for numerical values embedding~\cite{sun2018randomized,quote:karapiperis2017distance}. Here, we introduce the notion of Bit Vector mechanism.

    The BV (\textbf{B}it \textbf{V}ector) mechanism is first proposed for privacy-preserving record linkage~\cite{quote:karapiperis2017distance,karapiperis2018federal}. Given random variables $r = \{r_1, r_2, ..., r_s\}$, interval parameter $t$ and the length of data range $\mu = \lvert [L, U] \rvert$, the encoding process can be presented by such a series of hash functions:

    \begin{equation}
        h_{r_i}(x)= \left\{ \begin{array}{ll}
            1, & \quad x \in [r_i - t, r_i + t]\\
            0, & \quad \text{otherwise}\\
    \end{array} \right.
    \end{equation}

    With the BV encoding mechanism, the expected number $w$ of the set components which are set in each bit vector can be given by $\mathbb{E}[w]=s\cdot 2t / \mu$. As each scalar data shares the same expected $w$, BV mechanism provides indistinguishability, which can be used for privacy-preserving encoding. Also, it has been shown that the BV mechanism can preserve Euclidean distance in hamming space. Thus, it can be used for distance estimation in the anonymized space, for values with $d_E \le 2t$, the Euclidean distance can be estimated by Hamming distance in the anonymized space: $d_E =  \mu \cdot d_H / (2s)$. Based on these property, BV is used for privacy-preserving record linkage (PPRL).

\subsection{Local Differential Privacy}

    The concept of differential privacy is proposed by DWork in the context of statistical disclosure control~\cite{Dwork2006revised}. Recent researches have validated that mechanisms with differential privacy output accurate statistical information about the whole data while providing high privacy-preserving levels for single data in datasets. Based on differential privacy, the notion of LDP (\textbf{L}ocal \textbf{D}ifferential \textbf{P}rivacy) is also proposed to protect local privacy context from data analysis~\cite{kairouz2014extremal,wang2019collecting,wang2017locally,bassily2015local}. 

    \begin{definition}[Local Differential Privacy]
        A randomized algorithm $\mathcal{M}$ with domain $\mathbb{N}^{| \mathcal{X} |}$ is $(\epsilon,\delta)$-LDP if for all $\mathcal{S} \subseteq \text{Range} \mathcal{M}$ and for all $x,y$ in domain:
        \begin{equation}
        \Pr[\mathcal{M}(x) \in \mathcal{S}] \le e^{\epsilon}\Pr[\mathcal{M}(y) \in \mathcal{S}] + \delta
    \end{equation}
    \end{definition}

    Literately, RAPPOR~\cite{erlingsson2014rappor} was proposed for studying client data under the framework of differential privacy. In one-time RAPPOR, a value $v$ is first hashed into a bloom filter $B$ with length $k$ by a series of hash functions. Then permanent randomized response is added to $B$ to get $B^{'}$ before reported. However, the one-time RAPPOR mechanism can not be used for distance-aware encoding as the bloom filter is not distance-aware. To facilitate this, we use the BV mechanism as mentioned. Instead of using permanent randomized response, the 1Bit mechanism~\cite{ding2017collecting} is used to embed a numerical value. For data from $0$ to $m$, a numerical value is encoded to $1$ with probability $\frac{1}{e^\epsilon+1} + \frac{x}{m} \cdot \frac{e^\epsilon-1}{e^\epsilon+1}$.

    The 1Bit and RAPPOR mechanisms have made great progress in some statistical areas such as mean estimation~\cite{Ding2018ComparingPM} and frequency estimation~\cite{wang2018locally}. Intuitively, we want to achieve LDP in the BV mechanism to provide a privacy-preserving distance-aware encoding mechanism.

\subsection{Distributed clustering}

    In the distributed environment, each data owner performs generalized a noised version of his dataset and sends the perturbated datasets to aggregator for clustering (non-interactive mode). To provide privacy guarantee, obfuscation mechanisms such as perturbation and dimensionality-reduction methods are used for data anonymization. The additive data perturbation (ADP~\cite{Oliveira2003Privacy}) and the random subspace projection (RSP~\cite{liu2006random}) are two of the most common approaches to transfer original data to the anonymized space in the literature.

    \textbf{1) ADP} (\textbf{A}dditive \textbf{D}ata \textbf{P}erturbation): Each party generalizes a noisy database by adding independent and identically distributed Gaussian noises to records. each entry $p \in \mathcal{D}_i$ is replaced by $p^{'} = p + noise$ ($noise\sim N(0,\sigma)$). Usually, the noise levels represent the privacy-preserving level.

    \textbf{2) RSP} (\textbf{R}andom \textbf{S}ubspace \textbf{P}rojection): In the random subspace projection setting, the privacy of the source data is guaranteed as the projecting process is non-invertible. A $d$-dimensional data can be projected to a $q$-dimensional vector with a $d \times q$ random Gaussian matrix $R$ by mechanism $p^{'} = \frac{1}{\sqrt{q}\sigma}pR$. It has been shown that the RSP can preserve the Euclidean distance.

    The ADP and RSP based mechanism can be used for non interactive clustering. However, both of them lack a strict privacy-preserving guarantee of these methods. To achieve this, we use differential privacy in the anonymized output of Bit Vector mechanism.

\subsection{Notations}

    \begin{table}[ht]
    \begin{center}  
    \caption{Notations in our paper}  
    \label{table:notion}
    \begin{tabular}{|p{2.5cm}<{\centering}|p{3.5cm}<{\centering}|}
        \hline  
        Notations & Explanations\\  
        \hline
        $\mathcal{X}$ & Universe \\
        $\mathcal{D} = \{p_1, ..., p_n\}$ & Dataset \\  
        $[L, U]$ & Data range \\
        $t$ & BV parameters \\
        $\mathcal{M}(\cdot)$ & Encoding mechanism \\ $C_i(i \in {1,2,...,k})$ & Clusters \\ 
        $d_E(x,y)$ & Euclidean Distance \\
        $d_H(x,y)$ & Hamming Distance \\
        $\hat{d}_E(x,y), \hat{d}_H(x,y)$ & Estimated distance \\
        $\epsilon, \delta$ & Privacy parameters\\
        \hline 
    \end{tabular}  
    \end{center}  
    \end{table}

    Our paper focuses on non-interactive clustering using local differential privacy. Before we formulate the privacy-preserving clustering process across multiple data sources, notations used in this paper are defined in Table~\ref{table:notion}. For convenience, it is assumed that data in each dimension is in Euclidean space.

\section{Expanding Utilities and Privacy Guarantees of BV}
\label{section: Towards functionalities: Bit Vector Mechanism}

    \begin{figure}[t]
        \center
        \includegraphics[width=85mm]{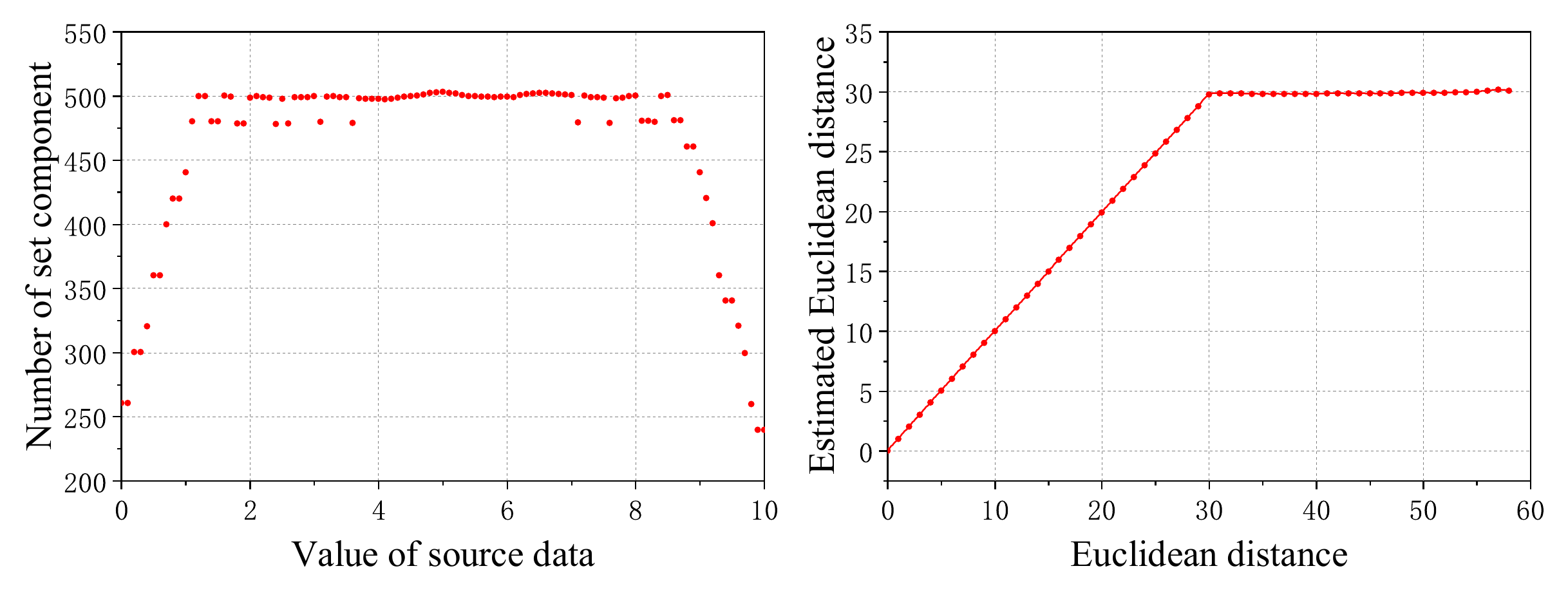}
        \caption{The drawback of BV mechanism. In the left example, data ranges in $[0,10]$ and $t=1.2$. In the right example, data ranges in $[0,100]$ and $2t=30$.}
    \label{fig:bv drawback}
    \end{figure}

    The BV mechanism has been shown capable for privacy preserving record linkage due to its distance-aware property. However, we found that this mechanism has some limitations when this mechanism is used in real life applications from the aspect of privacy protection and usabilities (Figure~\ref{fig:bv drawback}). 

    Firstly, the BV mechanism guarantees privacy from the perspective that the number of set components in different bit vectors stays the same statistically whatever the value is. Under such fact, an adversary can not retrieve the original data from received bit vectors without knowing random variables. However, our simulations show that values around $L$ or $U$ do not follow this rule. More seriously, the experiments show that the BV mechanism can only preserve distance in $2t$. When used in record linkage scenario, this property does not hurt much. however, when in clustering, this drawback would cause errors. 

    We fix the first problem by by extending $U$ to $U + t$ and $L$ to $L-t$, then modify $\mu$ from $\mu = U-L$ to $\mu=U-L + 2t$. This improvement is easy to be implemented in the BV mechanism and is included in this paper. For the limitations of usabilities, we propose a distance consistence algorithm for whole range distance estimation. To achieve rigorous privacy guarantee, we then introduce the differentially private bit vector mechanism and then analyze the decoding performance theoretically.

\subsection{Whole range distance estimation}

    We first show that even though the BV mechanism can only preserve Euclidean within a small range ($2t$ at most), we can still estimate distance over $2t$.

    \begin{figure}[h]
        \center
        \includegraphics[width=85mm]{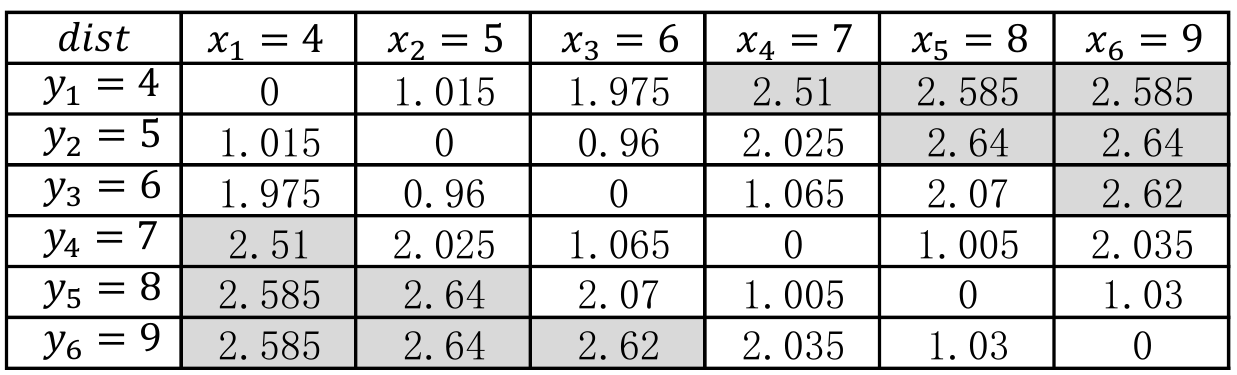}
        \caption{An example of wrong distance estimation. 1000 hash functions are used and the range of data is $[0,20]$.}
        \label{fig: bv wrong estimation}
    \end{figure}

    In Figure~\ref{fig: bv wrong estimation}, we show that when the true Euclidean distance exceeds $2t$, the estimation goes wrong. In this example, we set the interval parameter $t=1.2$ and the length of bit vector $s = 1000$ with data range in $[L, U] = [0, 20]$.  For example, the distance between $y_2 = 5$ and $x_6 = 9$ should be around $4$, not $2.16$. 

    % It is clearly that the estimation goes wrong when the true Euclidean distance is greater than $2t$. Also, we find that those wrong estimated distances are around $2t$.

    To solve this problem, we first define \textbf{\textit{local view}}, \textbf{\textit{global view}} and \textbf{\textit{unreachable}} distance. Then we find that the distance is consistent even in the anonymized space.

    \begin{definition}
        For data $x$ and $y$, we say that $(x,y)$ are in \textbf{\textit{local view}} iff $d_E(x,y) \le 2t$, and $(x,y)$ are in \textbf{\textit{global view}} iff $d_E>2t$ and there are limited values $v_1 \le v_2 \le ... \le v_k$, such that:
        
        \begin{equation}
            d_E(x,y) = d_E(x,v_1) + \sum_{i = 1}^{k-1} d_E(v_i, v_{i+1}) + d_E(v_k, y)
        \end{equation}
        Otherwise, we say that $x$ and $y$ are \textbf{\textit{unreachable}}.
    \end{definition}

    \begin{theorem}[Distance consistence]
    \label{theorem: distance consistence}
        For numerical values $x \le y \le z$ with both of them in local view, we have:
        
        \begin{equation}
            \hat{d}_E(x,z) = \hat{d}_E(x,y) + \hat{d}_E(y,z)
        \end{equation}
        
        \begin{proof}
            Let $ \tau =\hat{d}_E(x,y) + \hat{d}_E(y,z)-\hat{d}_E(x,z)$, thus, $\tau = \frac{u}{2s}[d_H(x,y) + d_H(y,z) - d_H(x,z)]$. For short, we use $\iota$ representing the bit is $0$ or $1$. As an example, when $\iota=0$, the triple $[\iota, \iota, \overline{\iota}]$ equals $[0,0,1]$ and the estimated Euclidean distance is in consistence. All the situation can be summarized in the following table:
        
            \begin{table}[ht]
            \begin{center}  
            \caption{Situations of bit value in bit vectors}  
            \label{table: situations of bit vector}
            \begin{tabular}{|p{1cm}<{\centering}|p{1cm}<{\centering}|p{1cm}<{\centering}|p{1cm}<{\centering}|p{2cm}<{\centering}|}  
                \hline  
                $bv(x)$  & $bv(y)$ & $bv(z)$ & $2s\tau /u$ & consistence\\  
                \hline
                $\iota$ & $\iota $ & $\iota$ & $0$ & \textit{true}\\
                $\iota$ & $\iota $ & $\overline{\iota}$ & $0$ & \textit{true}\\
                $\iota$ & $\overline{\iota} $ & $\iota$ & $2$ & \textit{false}(*)\\
                $\iota$ & $\overline{\iota} $ & $\overline{\iota}$ & $0$ & \textit{true}\\
                \hline 
            \end{tabular}  
            \end{center}  
            \end{table}
        
            With situations in Table~\ref{table: situations of bit vector} (*), we can find that only when the correspond bits in bit vector of $[x,y,z]$ equal to $[0,1,0]$ or $[1,0,1]$, the consistence fails. We will show that this is impossible. We first analyze the case of $[x,y,z] = [0,1,0]$. When $[rbv(x), rbv(y) = 0,1]$, it means that $x+t < r$, and when $[rbv(y), rbv(z) = 1,0]$, it means that $r < z-t$. This corresponds to the equation:
        
            \begin{equation}
                \left\{ \begin{array}{ll}
                x+t < r \\
                r < z-t \\
                \end{array} \right.
            \end{equation}
            
            Which means that $(z-t)-(x+t) = (z-x) - 2t > 0$. This conflicts with the assumption that $x,y,z$ are in \textbf{\textit{local view}}, which means that $z-x \le 2t$. Analogous to $[x,y,z] = [0,1,0]$, the situation of $[1,0,1]$ can also be proved unsatisfied. 
        \end{proof}
    \end{theorem}
    
    Just as the distance consistence in Euclidean space, we can adjust distance in global view with distances in local view. The pseudo-code of distance consistence algorithm using global view is described in Algorithm~\ref{algorithm: distance consistence}. As parameter $t$ is not revealed to the aggregator, we should find the range that holds local view (line 2). We first preserve distance in local view (line 3), then distance in global view are adjusted with the distance consistence theorem (lines 4-6). At last, the unreached distance are kept unchanged (line 7). In our implementation, a flag matrix recording in which iterations $D_{i,j}$ is revised is included, and the distance can only be updated with modified distance before current iterations (line 5).

    The distance consistence algorithm cannot be used to modify the unreachable distance. To solve this problem, as for the custodian, we recommend to add some mediate vales for embedding. For example, when $2t < 3$, the distance of $3.4$ and $7.9$ are unreachable. The data owner can then generate a noisy value $5.5$. In this way, the Euclidean distance can be estimated by $\hat{d}_E(3.4, 5.5) + \hat{d}_E(5.5, 7.9)$. It should be noticed adding external values increases computing complexity. 

    \begin{algorithm}[t]
    \caption{Distance consistence algorithm} 
    \label{algorithm: distance consistence}
    \begin{algorithmic}[1]
        \REQUIRE 
        A distance matrix $D$
        \STATE initialize a new distance matrix $\hat{D}$, $\forall i,j, \hat{D}_{i,j} = \infty$
        \STATE $r = \max \{D_{i,k} | \exists i,j,k :  D_{i,j} + D_{j,k} = D_{i,k}\}$
        \STATE $\forall D_{i,j} \le r$: $\hat{D}_{i,j}=D_{i,j}$
        \FOR {$\hat{D}_{i,j} \le r$ and $\hat{D}_{j,k} \le r$ and $\hat{D}_{i,k} = \infty$}
        \STATE update $\hat{D}_{i,k} = \hat{D}_{i,j} + \hat{D}_{j,k}$
        \ENDFOR
        \STATE $\forall \hat{D}_{i,j}=\infty$, $\hat{D}_{i,j} = D_{i,j}$
        \RETURN the refined distance matrix $\hat{D}$
    \end{algorithmic}
    \end{algorithm}

\subsection{Differentially Private Bit Vector Encoding}

    For the the single data encoding, the probability function of Bit Vector can be written as $\Pr [y=1] = 2t / \mu$. From the aspect of differential privacy, it provides $0$-DP, and no utilities are guaranteed. To make it feasible, the random variables are kept unchanged when generated, which means:

    \begin{equation}
    \label{equation: bv Posterior}
        \Pr [y=1|r_i, t]=\Pr [x \in [r_i-t, r_i +t]]
    \end{equation}

    In this way, the distance information is preserved in the Hamming space, because we have $\Pr[y_{ai}=1,y_{bi}=1|r_i, t] = \Pr[r_i-t\le x_a \le x_b \le r_i + t]$. It implies that $d_E(x_a, x_b) \le 2t$. In a honest-but-curious setting, only the distance information is known to the aggregator. However, this mechanism is not privacy-preserving with a malicious adversary or in the two-party setting. According to Equation~\ref{equation: bv Posterior}, the probability of $\Pr[x|r_i, t, y]$ can be learned. More importantly, according to the distance consistence algorithm, the possible $t$ can be estimated with the distance matrix. Under such assumptions, the BV mechanism is vulnerable under observation of $r_i$. Like RAPPOR mechanism, we use a 1Bit-like mechanism in each set bit. The probability function is:

    \begin{align}
    \label{equation: dpbv}
        \Pr[y=1|r_i, t] &= \frac{e^\epsilon}{e^\epsilon+1}\cdot \Pr[x \in [r_i-t, r_i +t]] \nonumber\\
        &+ \frac{1}{e^\epsilon+1} \cdot  \Pr[x \notin [r_i-t, r_i +t]]
    \end{align}

    In this paper, this encoding mechanism is called DPBV (\textbf{D}ifferentially \textbf{P}rivate \textbf{B}it \textbf{V}ector) mechanism. We will further show that the DPBV mechanism guarantees $(\epsilon, \delta)$-LDP and is distance-aware in the anonymized space.

    \begin{theorem}
        Encoding mechanism with Equation~\ref{equation: dpbv} achieves $\epsilon$-LDP.
        
        \begin{proof}
            In this mechanism, both $r_i$ and $t$ are kept unchanged when generated. The DPBV for single bit outputs $y=0$ or $y=1$ with probability of $\frac{1}{e^\epsilon+1}$ or $\frac{e^\epsilon}{e^\epsilon+1}$ (Equation~\ref{equation: dpbv}). Thus, for different numerical value $x_a, x_b \in [L, U]$ and any output $y$, we have:
            \begin{equation}
                \Pr[y|x_a,r_i, t] \le e^\epsilon \cdot \Pr[y|x_b, r_i, t]
            \end{equation}  
            Thus DPBV mechanism for one bit preserves $\epsilon$-LDP.
        \end{proof}
    \end{theorem}

    \begin{theorem}[Expected number of set components]
    \label{equation:expected common components}
        
        In the DPBV setting, the expected number $w$ of components which are set in each bit vector is:
        
        \begin{equation}
            \mathbb{E}[w] = s \cdot (\frac{2t}{\mu} \cdot \frac{e^\epsilon-1}{e^\epsilon+1} + \frac{1}{e^\epsilon+1})
        \end{equation}
    \end{theorem}

    Theorem~\ref{equation:expected common components} indicates that the expected common number of components of different source values is the same. For the aggregator who receives the encoded data, data in source databases are indistinguishable.

    \begin{theorem}[DPBV-Composition]
    \label{theorem: dpbv-composition}
        Given random variables $\textbf{r} = \{r_1, r_2, ..., r_s\}$, the randomized response with bit vector satisfies $(\epsilon,\delta)$-local differential privacy, where:
        \begin{equation}
            \delta = (\frac{e^\epsilon}{e^\epsilon+1})^s - e^\epsilon\cdot (\frac{1}{e^\epsilon+1})^s
        \end{equation}
    \end{theorem}

    Theorem~\ref{theorem: dpbv-composition} gives the lower bound of the privacy-preserving level.For the space reasons, the proof is given in the appendix. In the following experimental setting, $s$ is usually very large. For example, the DPBV mechanism is $(2, 7.5\times 10^{-56})$-LDP when $s = 1000$.

\subsection{Distance-aware Decoding}

    With the DPBV encoding mechanism, each original value is embedded into a vector in Hamming space. In this section, we focus on computing Euclidean distance information in Hamming space.

    \begin{theorem}[Euclidean Distance Estimation]
    \label{theorem: euclidean estimation}
        Given Hamming distance $d_H$ between embeded vectors $\mathcal{M}(x_1)$ and $\mathcal{M}(x_2)$, the Euclidean distance between numerical values $x_1,x_2 \in [L, U]$ can be estimated by:
        
        \begin{equation}
            \hat{d}_E(x_1, x_2) = \frac{\mu }{2s}\cdot (\frac{e^\epsilon+1}{e^\epsilon-1})^2 \cdot d_H-\frac{\mu e^\epsilon}{(e^\epsilon-1)^2}
        \end{equation}
        
        \begin{proof}
            As it is stated in the DPBV mechanism, the process of adding differential privacy to the BV mechanism can be thought as the randomized response process in the encoding mechanism: the bits in BV results are kept unchanged with probability $\frac{e^\epsilon}{e^\epsilon+1}$ and reversed with probability $\frac{1}{e^\epsilon+1}$.
            
            From the encoding process, the expected hamming distance can be estimated by:
            
            \begin{equation}
            \label{equation:expected hamming distance}
                \mathbb{E}[d_H] = 2s \cdot \frac{d_E}{\mu} \cdot \frac{e^{2\epsilon}+1}{(e^\epsilon+1)^2} + \big[s - 2s \cdot \frac{d_E}{\mu}\big] \cdot \frac{2e^\epsilon}{(e^\epsilon+1)^2} 
            \end{equation}

        \end{proof}
    \end{theorem}

    With the correlation between $d_E$ and $d_H$, we can then use $d_H$ in the anonymized space to estimate the Euclidean distance. We can also prove that the error of distance estimation is bounded (the proof is in the appendix).

    \begin{theorem}
        For value $x_1$, $x_2$ with $d_E = |x_1 - x_2|$, the aggregator can estimate the distance $\hat{d}_E$ with Theorem~\ref{theorem: euclidean estimation}. With probability at least $1-\beta$, we have:
        \begin{equation}
            |\hat{d}_E-d_E| \le \frac{\mu}{2} \cdot (\frac{e^\epsilon+1}{e^\epsilon-1})^2 \sqrt{\frac{\ln\frac{2}{\beta}}{2s}}
        \end{equation}
    \end{theorem}

\section{Clustering on anonymous data}
\label{section:clustering_with_anonymous_data}

    When consolidated by the aggregator, data are in the anonymous space. Analysis on the integrated anonymous dataset is limited because we can only estimate distance in the Hamming space. Motivated by the k-means algorithm, we now present the kCluster algorithm.

\subsection{KCluster clustering method}

    The k-means clustering algorithm~\cite{macqueen1967some} is one of the most fundamental clustering methods. It aims at partitioning all the data points into $k$ clusters by minimizing the within-cluster sum of squares (denote $u_i$ as the mean of points in cluster $C_i$):
    
    \begin{equation}
        \arg\min \sum_{i \in [k]} \sum_{p\in{C_i}} (p-u_i)^2 
    \end{equation}

    \begin{algorithm}[t]
    \caption{DP-kCluster: differentially private clustering} 
    \label{algorithm: Private-preserving Clustering with Bit Vectors}
    \begin{algorithmic}[1]
        \setlength{\textfloatsep}{5pt}
        \REQUIRE 
        The number of clusters, $k$; anonymized dataset $\mathcal{D}^{'}_{i\in [n]}$ from $n$ data providers; Decoding parameters, $s$ and $\mu$; privacy-preserving level $\epsilon$;
        \STATE $\mathcal{D}^{'} = \mathcal{D}^{'}_1 \cup \mathcal{D}^{'}_2 \cup ... \cup \mathcal{D}^{'}_n$
        \STATE Randomly choose $k$ records as initial Clusters $C_1, C_2, ..., C_k$.
        \STATE Assign each record into its nearest cluster.
        \STATE $l = |\mathcal{D}^{'}|$
        \REPEAT
        \STATE Generate clusters: $\forall i \le k$, $C_{i}^{'} = C_i$, $C_i = \varnothing$
        \FOR {$j = 1, 2, 3, ..., l$}
        \STATE $index = \mathop{\arg\min}\limits_{t \in \{1,2,...,k\}} \hat{D}_\mathcal{C}(\mathcal{M}(p_j), C_t^{'}) $
        \STATE $C_{index} = C_{index} \cup {\mathcal{M}(p_i)}$
        \ENDFOR
        \UNTIL{$\forall i \in \{1,2,...,k\},  C_i = C_i^{'}$}
        \RETURN Set of clusters $C = \{C_1, C_2, ..., C_k\}$
    \end{algorithmic}
    \end{algorithm}

    However, the DPBV mechanism is not suitable for k-means as calculating the mean value is not supported. Instead of assigning a point to its closest center, we assign a point to its closest cluster. Given a set of observations $\{p_1, p_2, ..., p_n\}$, we define the average distance between point $p$ and cluster $\mathcal{C}$ to be:

    \begin{equation}
    \label{equation:point to cluster distance}
        D_{\mathcal{C}}(p, C_i) = \frac{\sum_{p^{'} \in C_i} d_E(p, p^{'})}{|C_i|}
    \end{equation}

    With the anonymized data $\{\mathcal{M}(p_1), \mathcal{M}(p_2), ..., \mathcal{M}(p_n)\}$, the distance between an anonymized point and a cluster $D_{\mathcal{C}}$ can be estimated by:

    \begin{equation}
    \label{equation:estimate point to cluster distance}
        \hat{D}_{\mathcal{C}}(\mathcal{M}(p), C_i) = \frac{\mu \cdot \sum_{\mathcal{M}(p^{'}) \in C_i} \big[ d_H(\mathcal{M}(p), \mathcal{M}(p^{'})
        - \frac{2s \cdot e^\epsilon}{(e^\epsilon+1)^2} \big]}{2s \cdot (\frac{e^\epsilon-1}{e^\epsilon+1})^2 \cdot |C_i|} 
    \end{equation}

    Based on $D_{\mathcal{C}}$, the clustering result is given by finding the objective $C$:

    \begin{equation}
        \mathop{\arg\min}_{C_1, C_2, ..., C_k} \sum_{i \in [k]} \sum_{\mathcal{M}(p) \in C_i} D_\mathcal{C}(\mathcal{M}(p), C_i) 
    \end{equation}

    In the distributed environment, dataset are integrated and then the DP-kCluster algorithm is run with Algorithm~\ref{algorithm: Private-preserving Clustering with Bit Vectors}. To produce the final clustering result, kCluster uses iteration to get a refined result each step. There are two main steps in the kCluster algorithm.

    \begin{itemize}[leftmargin=*]
        \item \textbf{Step 1: Initializing $k$ clusters.} Choose $k$ points as the initial centroids (line 2-3). Form $k$ clusters by arranging each point to its nearest centroid. 
        \item \textbf{Step 2: Iteration.} In the $j$-th iteration, for each point $p$, find the closest cluster in $(j-1)$-th iteration and reset $p$'s label (line 5-11).
    \end{itemize}

    We will further show that the DPBV encoding mechanism can be used for anonymized clustering with current clustering algorithms. Take DBSCAN as a example. In DBSCAN clustering algorithm, given distance parameter $E$, one essential task is to find out the number of points within distance $E$. In Hamming space, the distance threshold is estimated by Equation~\ref{equation:expected hamming distance}. Also, the DPBV encoding mechanism can be used for hierarchical clustering.

\subsection{Decomposition for Vertically Partitioned Data}

    In this section we focus on calculating distances between records owned by distributed data custodians. It is different from the centralized setting that distance should be calculated on each side of data custodian. For convenience, we assume that data are held separately by \emph{Alice} and \emph{Bob}, and \emph{Alice} wants to know the Euclidean distance between record pair $(p_A, p_B$). The target is to compute:
    
    \begin{equation}
    \label{equation:distributed target}
        d_E(p_{A}, p_{B})^2 = \sum_{i = 1}^{d}(p_{A,i} - p_{B,i})^2 
    \end{equation}

    In the horizontally partitioned setting, data held by \emph{Alice} and \emph{Bob} need to be encoded into the Hamming space. The embedded data from \emph{Bob} are then sent to \emph{Alice}. From the side of \emph{Alice}, the distance can be estimated by:

    \begin{equation}
    \label{equ: horizontal distance estimation}
        \hat{d}_E(p_A, p_B) = \frac{1}{2s(\frac{e^\epsilon-1}{e^\epsilon+1})^2} \sqrt {\sum_{i = 1}^{d} \mu^2[d_H(p_{A,i}, p_{B,i})- \frac{2s \cdot e^\epsilon}{(e^\epsilon+1)^2}]^2} 
    \end{equation}

    \begin{figure}[t]
        \center
        \includegraphics[width=\linewidth]{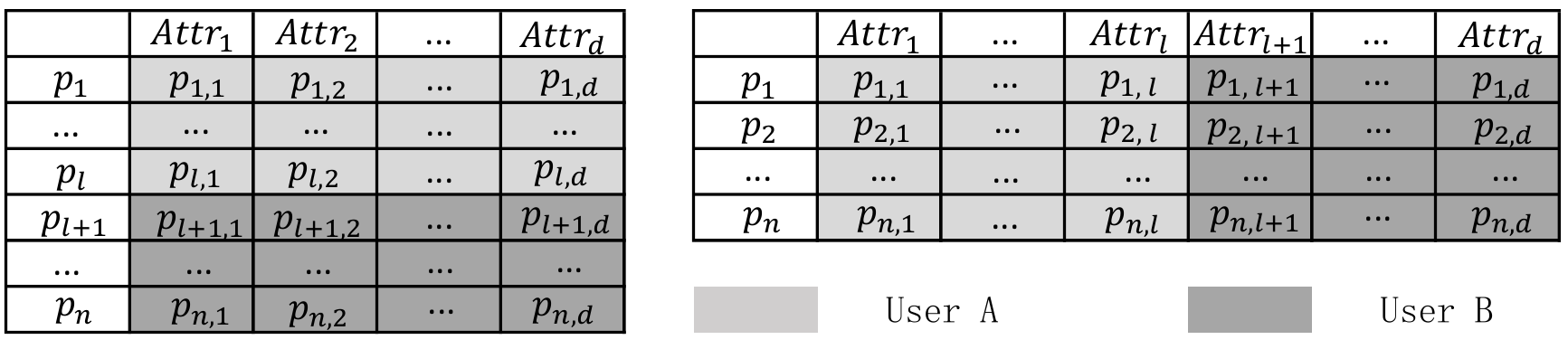}
        \caption{Horizontally partitioned data and vertically partitioned data.}
    \end{figure}

    In real life, data in distributed custodians may share common identifiers and different attributes (vertically partitioned data). Estimation distance in the vertical setting is different. We first define $L_{A,B}^2$ and $R_{A,B}^2$ as $L_{A,B}^2 = \sum_{i = 1}^{l} (p_{A,i}-p_{B,i})^2, R_{A,B}^2 = \sum_{i = l+1}^{d} (p_{A,i}-p_{B,i})^2$. For \emph{Alice}, $L_{A,B}^2$ can be calculated preciously without privacy leakage. One common way to estimated distance of \emph{Bob}'s part is to encode all of his data and then estimate the Euclidean distance by: 

    \begin{equation}
    \label{equation:vertical estimation method one}
        \hat{R}_{A,B}^2=  \frac{1}{4s^2(\frac{e^\epsilon-1}{e^\epsilon+1})^4}\big[\sum_{i=l+1}^{d} \mu^2[d_H(p_{A,i},p_{B,i})-\frac{2s\cdot e^\epsilon}{(e^\epsilon+1)^2}]\big]
    \end{equation}

    Considering that $\forall i \in [l+1, d]$, each value of $d_E(p_{A,i}, p_{B,i})$ can be calculated by \emph{Bob}, we think the errors of estimating $d_E(p_A, p_B)$ can be tightened.

    \begin{align}
        R_{A,B}^2 &= \sum_{i = l+1}^{d} (p_{A,i}-p_{B,i})^2 \\
        &= \sum_{i = l+1}^{d} (p_{A,i}^2 + p_{B,i}^2) - \sum_{i = l+1}^{d}{2 \cdot p_{A,i} \cdot p_{B,i}}
    \end{align}

    Let $\mu_{\max} = 2 \times \sum_{i=l+1}^{d} \mu^2$, then $R_{A,B}^2$ can be estimated by:

    \begin{align}
    \label{equation:vertical estimation method two}
    \hat{R}_{A,B}^2 &= \frac{\mu_{\max}}{2s(\frac{e^\epsilon-1}{e^\epsilon+1})^2} \big[ d_H[\sum_{i = l+1}^d p_{A,i}^2 + p_{B,i}^2, (\sum_{i = l+1}^d 2 \cdot p_{A,i} \cdot p_{B,i})] \nonumber\\
    &- s\cdot \frac{2e^\epsilon}{(e^\epsilon+1)^2}\big]
    \end{align}

    Compared with equation~\ref{equation:vertical estimation method one}, equation~\ref{equation:vertical estimation method two} only needs one-time encoding for Euclidean distance estimation. In both situation, the distance between record pair $(p_A, p_B)$ can be estimated by $\hat{d}_E(p_A,p_B) = \sqrt{L_{A,B}^2 + \hat{R}_{A,B}^2}$. In the experimental part, we will analyze errors of these two distance estimation methods.

\section{Experiments}
\label{section: experiments}

    In this section, we first measure the distance consistence in the anonymized space, then the decomposition for vertically partitioned data are analyzed. At last, clustering performance with existing algorithms are presented.

    \textbf{Datasets.} We choose different types of data to evaluate our clustering algorithms. In the visualization part, we use three publicly available datasets: blobs based (Aggregation dataset~\cite{gionis2007clustering}),circles based (pathbased dataset~\cite{chang2008robust}) and moon-shape based dataset (``twomoons''~\cite{rozza2014novel}).  We also use a real life dataset: the digit dataset~\cite{scikit-learn}, composed of 1797 images, each image is a $8 \times 8$ hand-written digit.

    \textbf{Parameter selection.} We embed each numerical value into Hamming space with $s = 1000$. For demonstration purpose, data are regularized to $[0, 50]$ in our experiments. The interval parameter is set $t=25$ in demonstration.

    \textbf{Methodology.} We choose kCluster and DBSCAN as basic clustering algorithms. We first embed source data with BV and DPBV mechanism, then we retrieve the distance matrix and use it for clustering. We use the Normalized Mutual Information to measure the clustering results. For comparison, we also cluster on the original dataset.

\subsection{Distance estimation utilities}
    
    \begin{figure}[ht]
    \center
    \includegraphics[width=0.9 \linewidth]{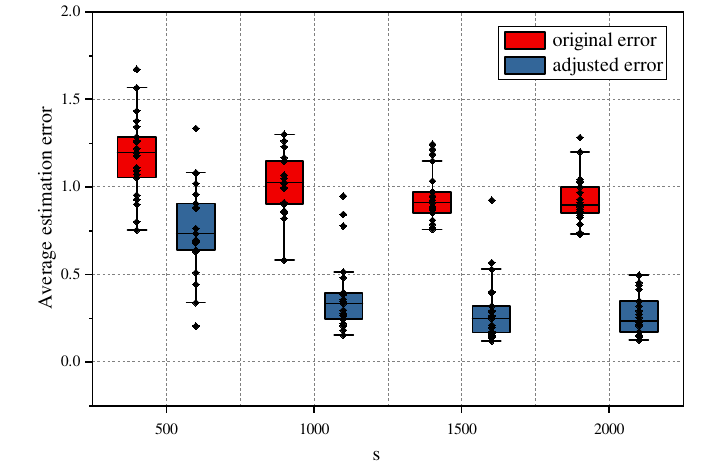}
    \caption{Estimation error with s changes.}
    \label{fig:experiment-distance-consistence}
    \end{figure}

    First, we implement the distance consistence algorithm and evaluate the average estimation error. A uniformly distributed dataset within range $[0, 25]$ is generated and encoded with $t=3$. Each time $10,000$ pairs are compared. The average estimation error is given by:

    \begin{equation}
    \label{equation:average error}
        ERR_{avg} = \frac{1}{|S_x||S_y|}\sum_{x\in S_x, y \in S_y} |d_E(x,y)-\hat{d}_E(x,y)|
    \end{equation}

    We can see from Figure~\ref{fig:experiment-distance-consistence} that with $s$ grows, the average error given by Equation~\ref{equation:average error} decreases. With a fixed $s$, we can conclude that the distance consistence algorithm can improve the performance of distance estimation.

\subsection{Partition analysis}

    In this part, we consider distance estimation over multidimensional data. The error of horizontally partitioned setting are not covered as because it is the same as the non-decomposition method in our experiment. For convenience, we set the same dimension of different data custodians. When encoding with non-decomposition method, each record is encoded $d$ times with DPBV mechanism, while it only cost two times for the Decomposition. As we know, the range of encoded data expands when decomposition, encoding with $s$ random variables would bring extra errors, thus the number of random variables we use in Decomposition is the same as that of non-decomposition.

    \begin{figure}[ht]
        \center
        \includegraphics[width=0.9 \linewidth]{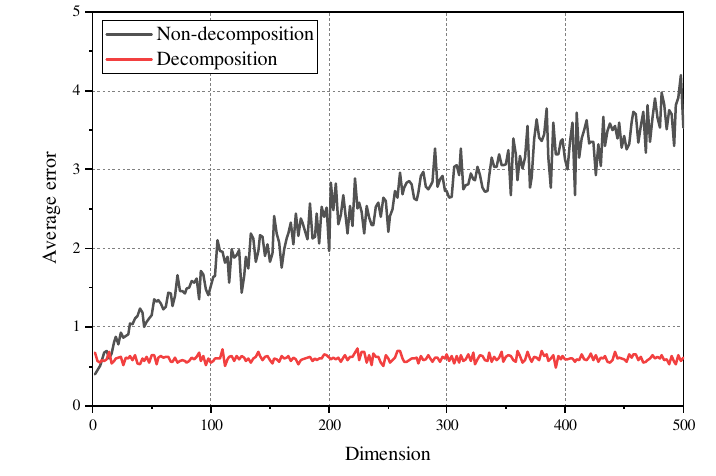}
        \caption{Average distance estimation error for vertically partitioned data.}
        \label{fig: decomposition}
    \end{figure}

    From Figure~\ref{fig: decomposition}, it is clear that with the increasing of data dimensions, the average error becomes larger. The main reason is that with dimensions increases, the errors accumulate with the times of encoding. As for Decomposition method, it only encodes two time whatever dimension is, no encoding error is contained, thus the error is controlled with explosion of dimensions.

\subsection{Clustering performances}

    \begin{figure*}[ht]
        \center
        \includegraphics[width=0.9 \linewidth]{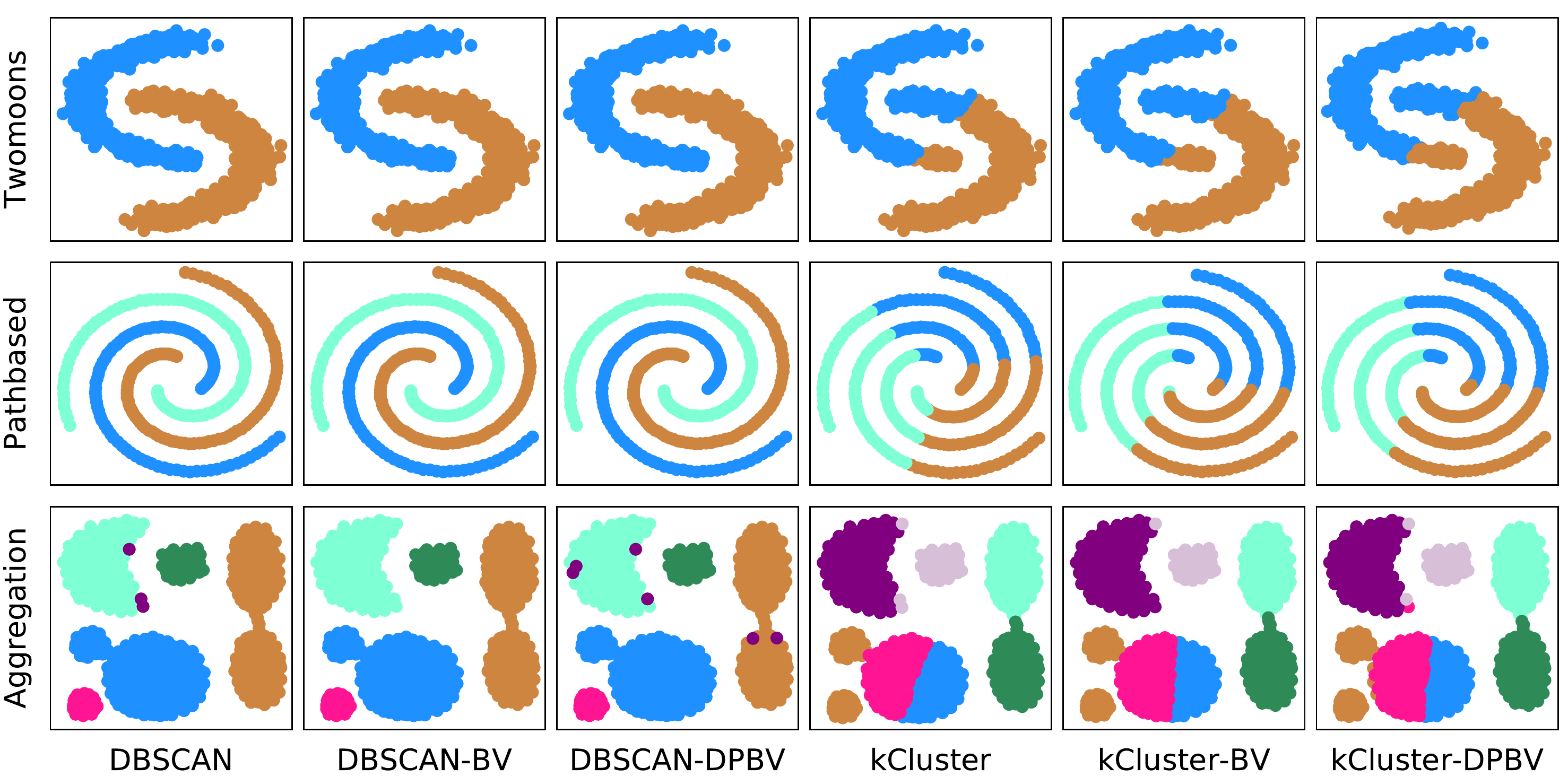}
        \caption{Visualization of different clustering algorithms.}
        \label{fig:clustering visualization}
    \end{figure*}

    In this section, we compare proposed algorithms with existing methods. Firstly, the visualization of mentioned clustering algorithm is shown in Figure~\ref{fig:clustering visualization}. We can see that the clustering results are not highly affected by anonymization. The results of privacy-preserving clustering algorithms are not exactly the same as original ones because distance estimation between two points is probabilistic. 

    \newcommand{\tabincell}[2]{\begin{tabular}{@{}#1@{}}#2\end{tabular}}
    \begin{table}[ht]
    \begin{center}
    \caption{clustering results}
    \label{table:high-dimensional data clustering}
    \setlength{\tabcolsep}{4mm}{
    \begin{tabular}{ccc}
        \hline
        clustering methods & privacy level& NMI \\
        \hline
        k-means & - & $74.32\%$ \\
        \hline
        RSP+k-means& \tabincell{c}{$50\%$\\$75\%$} &\tabincell{c}{$63.65\%$ \\$67.08\%$ }\\
        \hline 
        ADP+k-means & \tabincell{c}{$\sigma = 1$\\$\sigma=2$} & \tabincell{c}{$73.99\%$\\$72.72\%$} \\
        \hline 
        kCluster & -& $74.65\%$\\
        \hline 
        LDP+kCluster &\tabincell{c}{$(1,8.9\times 10^{-137})$-LDP\\$(2,7.5\times 10^{-56})$-LDP} & \tabincell{c}{$70.89\%$\\$73.57\%$} \\
        \hline
    \end{tabular}}
    \end{center}
    \end{table}

    To better comprehend the impact of applying anonymization in clustering process. We run a series of experiments on the digit dataset. Each picture is transformed into a vector with length 64. We compare our privacy-preserving clustering algorithm with the RSP based and ADP based methods. For the ADP based algorithm, we keep the variance of noise $\sigma=1$ and $2$. For the RSP based method, we project its dimension to $50\%$ and $75\%$ of the original dimension. Then the transformed data are clustered using typical k-means algorithm. The clustering results are listed in Table~\ref{table:high-dimensional data clustering}. The performance of kCluster is better that that of k-means. Unfortunately, there lacks a baseline for comparing privacy-preserving level between $(\epsilon, \delta)$-LDP, ADP and RSP based clustering algorithms. While it should mention that our LDP is in the anonymization space, which can preserve semantic information. For ADP based mechanism, adding noise with $\sigma\in\{1,2\}$ can achieve high utilities, However, the range of data can be quite determinated after ADP when $\sigma$ is at a low level. For example, encoding value $x=1$ with $\sigma=2$, we get $3$. From the perturbated value we can still be sure with high confidence that the original data is not big. From this perspective, using LDP in the anonymized space preserves higher privacy-preserving level.

    \begin{figure}[!h]
        \center
        \includegraphics[width=0.9\linewidth]{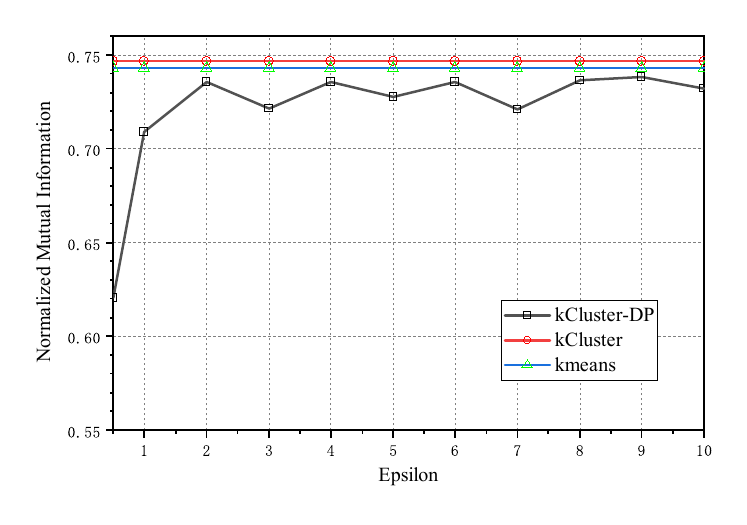}
        \caption{Performance of DP-kCluster with epsilon.}
        \label{fig: epsilon to clustering}
    \end{figure}

    Shown in Figure~\ref{fig: epsilon to clustering}, we also test the influence of $\epsilon$ to the clustering results (with $\delta = (\frac{e^\epsilon}{e^\epsilon+1})^s - e^\epsilon\cdot (\frac{1}{e^\epsilon+1})^s$). As the encoding process is randomized, the clustering performance fluctuates within a small range. According our experiments, we can achieve high utility with $\epsilon \ge 1$. Under such configuration, high privacy-preserving level is guaranteed.

\section{Privacy analysis and limitations}
\label{section:privacy analysis}

    \begin{figure}[ht]
        \center
        \includegraphics[width=0.9\linewidth]{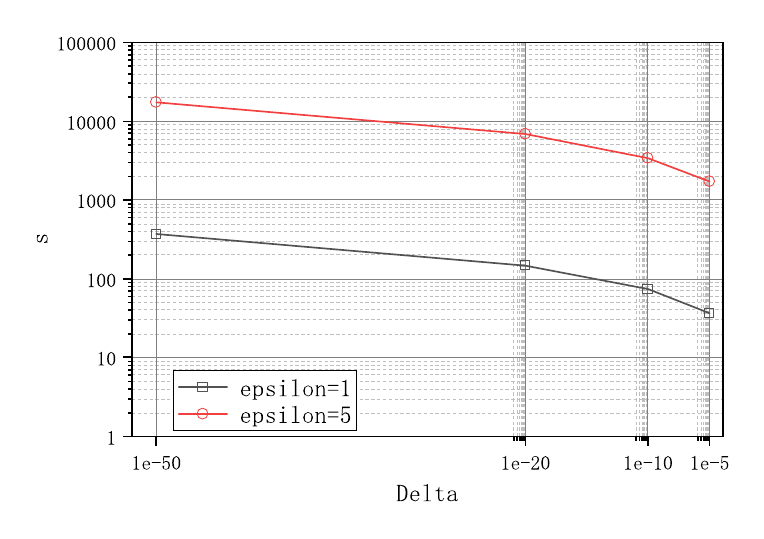}
        \caption{the length of $s$ under $(\epsilon, \delta)$-LDP.}
        \label{fig: s with epsilon and delta}
    \end{figure}

    As detailed in Section~\ref{section: Towards functionalities: Bit Vector Mechanism}, with an anonymization mechanism (BV), each scalar value is turned into a bit vector. Then with the guarantee of LDP, each vector in $\{0, 1\}^{s}$ is noised after perturbation. In this way, the privacy is guaranteed by the anonymization process and the perturbation process. To achieve $(\epsilon, \delta)$-LDP, we can set $s$ with $s = \lceil \frac{\ln \delta}{\epsilon - \ln (e^\epsilon+1)}\rceil$. Show in Figure~\ref{fig: s with epsilon and delta}, with the increase of $\epsilon$, the length of anonymized bit vector increases. Also, we notice that $s$ also increases when $\delta$ decreases. Technologically, this is because we expand the anonymized space to achieve lower $\delta$, which means that $\Pr[\mathcal{M}(x)]$ descends.
    
    \begin{figure}[ht]
        \center
        \includegraphics[width=0.9\linewidth]{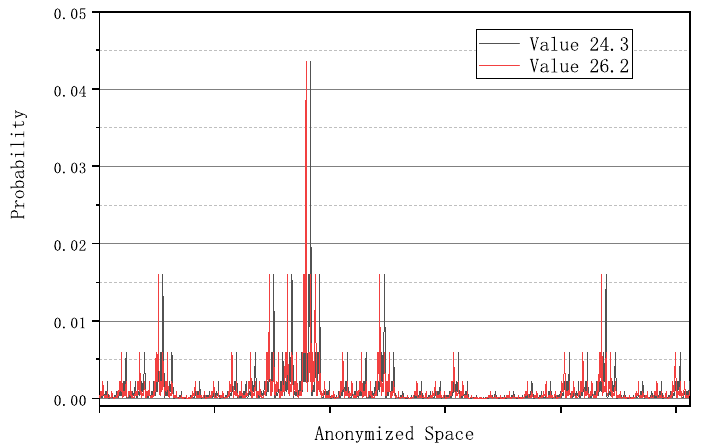}
        \caption{\textit{HBC} aggregator.}
        \label{fig:probability in the anonymized space}
    \end{figure}

    \begin{figure}[ht]
        \center
        \includegraphics[width=0.9\linewidth]{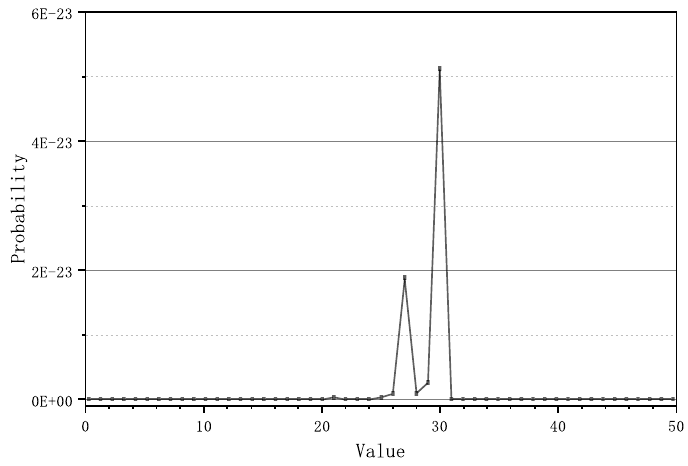}
        \caption{malicious adversary}
        \label{fig:fig_probability_space_source_value}
    \end{figure}

    We also consider two types of attackers, the \textit{Honest But Curious (HBC)} aggregator and a malicious attacker with access to all of the configuration parameters, including the data of range, random variables and $\epsilon$. For an \textit{HBC} aggregator. We can plot the probability of all possible output $o$ with input $x = 24.3$ and $x=26.2$ in the anonymized (Figure~\ref{fig:probability in the anonymized space}) under $\epsilon=1$. The probability space of value $24.3$ is likely to that of $26.2$. For demonstration, we only set $s=10$, as the anonymized space would be too large to be shown when $s$ becomes large. In this example, the privacy-preserving level is guaranteed by anonymized mechanism and LDP. For output $o \in \{0,1\}^s$, $\Pr[\mathcal{M}(x)=o]$ is very small.

    For the malicious attacker, we wonder if he can retrieve the original data. In this experiment, we set $s = 1000$ and $\epsilon = 1$. We can see in Figure~\ref{fig:fig_probability_space_source_value} that for each data $x \in \mathcal{X}$, we have $\Pr[\mathcal{M}(x)=o] \rightarrow 0$, which leads to indistinguishabilities. In this example, with received vector, the adversary might guess the source value to be in $[26,30]$. However, the DPBV mechanism cannot prevent collusion attacks. For the adversaries knowing original data and its corresponding bit vectors, representing know $\mathcal{X}_A$ and $\{DPBV(x) | \forall x \in \mathcal{X}_A\}$. To a certain extent, the value of $y$ can be retrieved when $s$ is large. The malicious adversary can retrieve $\hat{y}$ by $y \in \{x_a + d_E(x_a, y), x_a + d_E(x_a, y)\}$ as the distance information contains original data and we designed DPBV mechanism to be distance-aware.

\section{Conclusion and Discussion}
\label{section:conclusion}

    This work investigates encoding mechanism with LDP guarantees and its application in distributed clustering. Our results show validate that we can achieve $(\epsilon, \delta)$-local differential privacy guarantees in the anonymized space as well as high distance estimation and clustering utilities. Our proposed solution can be used in privacy-preserving data sharing and multi-party clustering in the distributed environment.

    As an application case, we designed a clustering algorithm for distributed clustering with only distance information in the anonymized space. A natural problem is that can this encoding mechanism be used in other analyzing tasks. As for future work, we plan to use DPBV mechanism for more aggregate statistics, such as mean estimation. We also wants this mechanism to be used in privacy-preserving classification.

    As far as we know, current methods with $\epsilon$-DP guarantees can only work in interactive clustering or clustering with a trusted aggregator. In this paper, we only use $(\epsilon, \delta)$-locally differentially private anonymization for data collection and analyzing. It stays an open question that can $\epsilon$-LDP be achieved in a anonymization mechanism that is still distance-aware? Up to now, we think it remains a challenge.

\bibliographystyle{IEEEtran}
\bibliography{IEEEquote}

\appendix
\section{Proof for DPBV composition}

    Given random variables $\textbf{r} = \{r_1, r_2, ..., r_s\}$, the randomized response with bit vector satisfies $(\epsilon,\delta)$-local differential privacy, where $\delta = (\frac{e^\epsilon}{e^\epsilon+1})^s - e^\epsilon\cdot (\frac{1}{e^\epsilon+1})^s$.
    
    \begin{proof}
        As random variables are given, $\forall r_i \in \textbf{r}$, we have:
        
        \begin{align}
            \Pr[B^{'} = b^{'} | X = x] &= \Pr[B^{'} = b^{'}|B=b,X=x] \cdot \Pr[B=b|X=x]  \nonumber\\ 
            &= \Pr[B^{'}=b^{'}|B=b]
        \end{align}
    
        First, we consider the situation on encoding with one bit ($s = 1$). Without loss of generality, for $b_i \in \{0,1\}$ in the bit vector encoding process, it holds that: 
        
        \begin{equation}
            \Pr[B_i^{'} = b_i^{'}|B_i = b_i] = (\frac{e^\epsilon}{e^\epsilon+1})^{b_i^{'}\odot b_i} \cdot (\frac{1}{e^\epsilon+1})^{1-b_i^{'}\odot b_i}
        \end{equation}
    
        The $b_i^{'}\odot b_i$ operation returns $1$ if $b_i^{'} = b_i$ and $0$ otherwise. Taking all random variables $\{r_1, r_2, ..., r_s\}$ into consideration, we have:
        
        \begin{align}
            \Pr[B^{'}&=b^{'}|B=b] = (\frac{e^\epsilon}{e^\epsilon+1})^{b_1^{'} \odot b_1} \cdot (\frac{1}{e^\epsilon+1})^{1-b_1^{'}\odot b_1} \times ...  \nonumber\\
            & \quad \quad \quad \quad \quad \times (\frac{e^\epsilon}{e^\epsilon+1})^{b_s^{'}\odot b_s} \cdot (\frac{1}{e^\epsilon+1})^{1-b_s^{'}\odot b_s} \nonumber\\
            & = (\frac{e^\epsilon}{e^\epsilon+1})^{\sum_{i = 1}^{s} (b_i^{'} \odot b_i)} \cdot (\frac{1}{e^\epsilon+1})^{s - \sum_{i = 1}^{s} (b_i^{'} \odot b_i)}
        \end{align}
    
        In this way, it is clear that $\Pr[B^{'}|B] \le (\frac{e^\epsilon}{e^\epsilon+1})^s = \Pr[B^{'}=B]$. And then it holds that, given $\forall B_A, B_B \in \{0,1\}^s$, $B_A \not= B_B$:
        
        \begin{equation}
            \Pr[B^{'}|B_A] \le e^\epsilon \Pr[B^{'}|B_B] + (\frac{e^\epsilon}{e^\epsilon+1})^s - e^\epsilon\cdot (\frac{1}{e^\epsilon+1})^s \nonumber
        \end{equation}  
    
        Thus, the DPBV achieves $(\epsilon, \delta)$-LDP, where $\delta = (\frac{e^\epsilon}{e^\epsilon+1})^s - e^\epsilon\cdot (\frac{1}{e^\epsilon+1})^s$.
    \end{proof}

\section{Proof for error bound of distance estimation}

    For value $x_1$, $x_2$ with $d_E = |x_1 - x_2|$, the aggregator can estimate the distance $\hat{d}_E$ with Lemma~\ref{theorem: euclidean estimation}. With probability at least $1-\beta$, we have:
    
    \begin{equation}
        |\hat{d}_E-d_E| \le \frac{\mu}{2} \cdot (\frac{e^\epsilon+1}{e^\epsilon-1})^2 \sqrt{\frac{\ln\frac{2}{\beta}}{2s}}
    \end{equation}

    \begin{proof}
        According to the Chernoff-Hoeffding bound~\cite{ding2017collecting}, we have:
        
        \begin{equation}
            \Pr\big[|d_H - \mathbb{E}[d_H]|\ge t\big] \le 2 \cdot e^{-\frac{2t^2}{s}}
        \end{equation}
        
        Then we get:
        \begin{equation}
            \Pr \big[| f^2 \cdot 2s \cdot \frac{\hat{d}_E}{\mu} - (\frac{e^\epsilon-1}{e^\epsilon+1})^2\cdot 2s \cdot \frac{d_E}{\mu}| \ge t \big] \le 2 \cdot e^{-\frac{2t^2}{s}}
        \end{equation}
        
        Which means:
        
        \begin{equation}
            \Pr \big[| \hat{d}_E - d_E| \ge \frac{ \mu t}{2s \cdot (\frac{e^\epsilon-1}{e^\epsilon+1})^2} \big] \le 2 \cdot e^{-\frac{2t^2}{s}}
        \end{equation}
        
        Thus by setting $t = \theta \cdot 2s \cdot (\frac{e^\epsilon-1}{e^\epsilon+1})^2$, we obtain:
        
        \begin{equation}
            \Pr[ |\hat{d}_E -d_E | \ge \theta \mu] \le 2 \cdot e^{-8 (\frac{e^\epsilon-1}{e^\epsilon+1})^4 s \theta^2}
        \end{equation}
        
        Set $\beta = 2 \cdot e^{-8 (\frac{e^\epsilon-1}{e^\epsilon+1})^4 s \theta^2}$, then the error is:
        
        \begin{equation}
            \theta \mu \le \frac{\mu}{2} \cdot (\frac{e^\epsilon+1}{e^\epsilon-1})^2 \sqrt{\frac{\ln\frac{2}{\beta}}{2s}}
        \end{equation}
        
        Thus, the proof is concluded
    \end{proof}

\end{document}